\documentclass{PoS}

\def\d{{\delta}}
\def\D{{\Delta}}
\def\e{{\epsilon}}
\def\g{{\gamma}}

\def\l{{\lambda}}

\def\O{{\Omega}}

\def\t{{\tau}}

\def\eqref#1{{(\ref{#1})}}
\def\eq#1{{Eq.~(\ref{#1})}}
\def\sec#1{{Sec.~\ref{#1}}}
\def\tab#1{{Table~\ref{#1}}}
\def\fig#1{{Figure~\ref{#1}}}

\usepackage{subcaption}
\usepackage{adjustbox}
\DeclareCaptionFont{CaptionFontSize}{\fontsize{10pt}{10pt}\selectfont}
\title{Lattice QCD Determination of $g_A$}

\ShortTitle{$g_A$ from QCD}

\author{\footnotesize{\speaker{Andr\'{e} Walker-Loud}
    {\normalfont , Lawrence Berkeley National Laboratory}}}

\author{\footnotesize{Evan Berkowitz
    {\normalfont , University of Maryland}}}

\author{\footnotesize{David A. Brantley, Arjun Gambhir, Pavlos Vranas
    {\normalfont , Lawrence Livermore National Laboratory}}}

\author{\footnotesize{Chris Bouchard
    {\normalfont , University of Glasgow}}}

\author{\footnotesize{Chia Cheng Chang
    {\normalfont , RIKEN-iTHEMS}}}

\author{\footnotesize{M.A.~Clark
    {\normalfont , NVIDIA Corporation}}}

\author{\footnotesize{Nicolas Garron
    {\normalfont , Liverpool Hope University}}}

\author{\footnotesize{B\'alint Jo\'o
    {\normalfont , Thomas Jefferson National Accelerator Facility}}}

\author{\footnotesize{Thorsten Kurth
    {\normalfont NERSC, Lawrence Berkeley National Laboratory}}}

\author{\footnotesize{Henry Monge-Camacho, Amy Nicholson
    {\normalfont , University of North Carolina Chapel Hill}}}

\author{\footnotesize{Christopher J Monahan, Kostas Orginos
    {\normalfont , The College of William \& Mary}}}

\author{\footnotesize{Enrico Rinaldi
    {\normalfont , Arithmer Inc. \& RIKEN-iTHEMS}}}

\abstract{
\phantom{space} The nucleon axial coupling, $g_A$, is a fundamental property of protons and neutrons, dictating the strength with which the weak axial current of the Standard Model couples to nucleons, and hence, the lifetime of a free neutron.  The prominence of $g_A$ in nuclear physics has made it a benchmark quantity with which to calibrate lattice QCD calculations of nucleon structure and more complex calculations of electroweak matrix elements in one and few nucleon systems.  There were a number of significant challenges in determining $g_A$, notably the notorious exponentially-bad signal-to-noise problem and the requirement for hundreds of thousands of stochastic samples, that rendered this goal more difficult to obtain than originally thought.

\phantom{space} I will describe the use of an unconventional computation method, coupled with ``ludicrously'' fast GPU code, access to publicly available lattice QCD configurations from MILC and access to leadership computing that have allowed these challenges to be overcome resulting in a determination of $g_A$ with 1\% precision and all sources of systematic uncertainty controlled.  I will discuss the implications of these results for the convergence of $SU(2)$ Chiral Perturbation theory for nucleons, as well as prospects for further improvements to $g_A$ (sub-percent precision, for which we have preliminary results) which is part of a more comprehensive application of lattice QCD to nuclear physics.
This is particularly exciting in light of the new CORAL supercomputers coming online, Sierra and Summit, for which our lattice QCD codes achieve a machine-to-machine speed up over Titan of an order of magnitude.
}

\FullConference{The 9th International workshop on Chiral Dynamics\\
		17-21 September 2018\\
		Durham, NC, USA}

\begin{document}

\section{Motivation \label{sec:intro}}
The nucleon axial coupling, $g_A$, often called the nucleon axial charge, is a ubiquitous quantity in nuclear physics.
The strength of this coupling controls the rate of nuclear reactions, beta-decay, and the pion-exchange contributions to the nucleon-nucleon potential.
Furthermore, it governs the lifetime of the free neutron and strongly influences the primordial abundances of H and $^4$He.

While the nucleon axial coupling has been measured extremely precisely experimentally, yielding a global average value of $g_A = 1.2732(23)$~\cite{PhysRevD.98.030001} (and updated results that are substantially more precise~\cite{Markisch:2018ndu}), first-principles calculations of this quantity provide a stringent test of the limits of the Standard Model (SM) and could potentially point to new physics. For example, the so-called ``neutron lifetime puzzle" (for a discussion, see Ref~\cite{Czarnecki:2018okw}), a 4-sigma discrepancy between experimental measurements utilizing beams of neutrons versus those using trapped ultracold neutrons, could point to new beyond the SM decay modes. A theoretical calculation rooted in the SM may help to clarify this puzzle. Furthermore, because it is so well-measured and ubiquitous, $g_A$ provides an important benchmark for lattice QCD (LQCD) calculations related to nuclear physics. Systematic errors of this quantity must be fully understood and controlled before calculations of more challenging, and less experimentally well-known, quantities, such as the axial form factor of the nucleon, can be regarded as reliable.

In these proceedings I will discuss the advances made that have enabled a recent calculation of $g_A$ to 1\% precision, with all systematics controlled. I will also discuss the implications of the results for the use of $SU(2)$ Chiral Perturbation theory for nucleons. Finally, I will present a preliminary sub-precision update of our results with improved statistics at the physical pion mass achieved using early science time on Sierra at LLNL, and discuss future prospects for further reduction of uncertainties.
This is an exciting time as we move towards the exascale era in which we aim to build a quantitative bridge between QCD and theories of nuclear physics, with the aim of building a predictive theory of nuclear structure and reactions, rooted in the Standard Model~\cite{Drischler:2019xuo}, beginning with properties of the nucleon, moving to light nuclei (see CD2018 talk of Z. Davoudi) and coupling to theories of many body nuclear physics (see the CD2018 talks of M. Piarulli and S. Pastore).

\section{A percent-level determination of $g_A$  from QCD\label{sec:gA}}
We have recently determined $g_A$ with an unprecedented percent-level of uncertainty~\cite{Chang:2018uxx}
\begin{equation}\label{eq:gA_callat}
g_A = 1.2711(103)^s(39)^\chi(15)^a(04)^V(55)^M\, .
\end{equation}
The sources of uncertainty are statistical ($s$), extrapolation to the physical pion mass ($\chi$), continuum extrapolation ($a$), infinite volume extrapolation ($V$) and a model average uncertainty ($M$).
Prior to this result, it was estimated that a 2\% uncertainty could be achieved with near-exascale computing (such as Summit at OLCF) by 2020~\cite{usqcd_doe_2016}.
There were several key features of our calculation that enabled a determination with 1\% uncertainty with the previous generation of supercomputers:
\begin{enumerate}
\item The use of an unconventional strategy motivated by the Feynman-Hellmann Theorem (FHT)~\cite{Bouchard:2016heu};

\item Access to publicly available configurations that enabled the full physical point extrapolation, in this case the $N_f=2+1+1$ HISQ~\cite{Follana:2006rc} configurations generated by MILC~\cite{Bazavov:2012xda};

\item \textit{Ludicrously} fast GPU code for lattice QCD, in this case the QUDA library~\cite{Clark:2009wm,Babich:2011np};

\item Access to leadership class computing at LLNL through the Grand Challenge Program and Titan at OLCF through the DOE INCITE program.

\end{enumerate}
I will describe the unconventional method in Sec.~\ref{sec:fh} and describe the comprehensive analysis that leads to the quoted uncertainty breakdown in Sec.~\ref{sec:extrap}, including the stability of the final extrapolation.  Implications for $SU(2)$ baryon $\chi$PT will be discussed in Sec.~\ref{sec:baryonChiPT} followed by a discussion of expected improvements in precision with some preliminary results in Sec.~\ref{sec:future}.

\subsection{An unconventional method \label{sec:fh}}
The two most pressing challenges in applying LQCD to nucleon elastic structure calculations are the exponentially bad signal-to-noise (S/N) problem~\cite{Lepage:1989hd} and contamination from excited states.
To overcome these challenges, we ``invented'' a new method for performing the calculations~\cite{Bouchard:2016heu} (after ``inventing'' this method, of course we realized it has been around since the 80's~\cite{Maiani:1987by,Gusken:1989ad} and there are more recent applications that are very similar~\cite{Bulava:2011yz,deDivitiis:2012vs,Chambers:2014qaa,Chambers:2015bka,Savage:2016kon}).

\begin{figure}[t]
\includegraphics[width=0.45\textwidth]{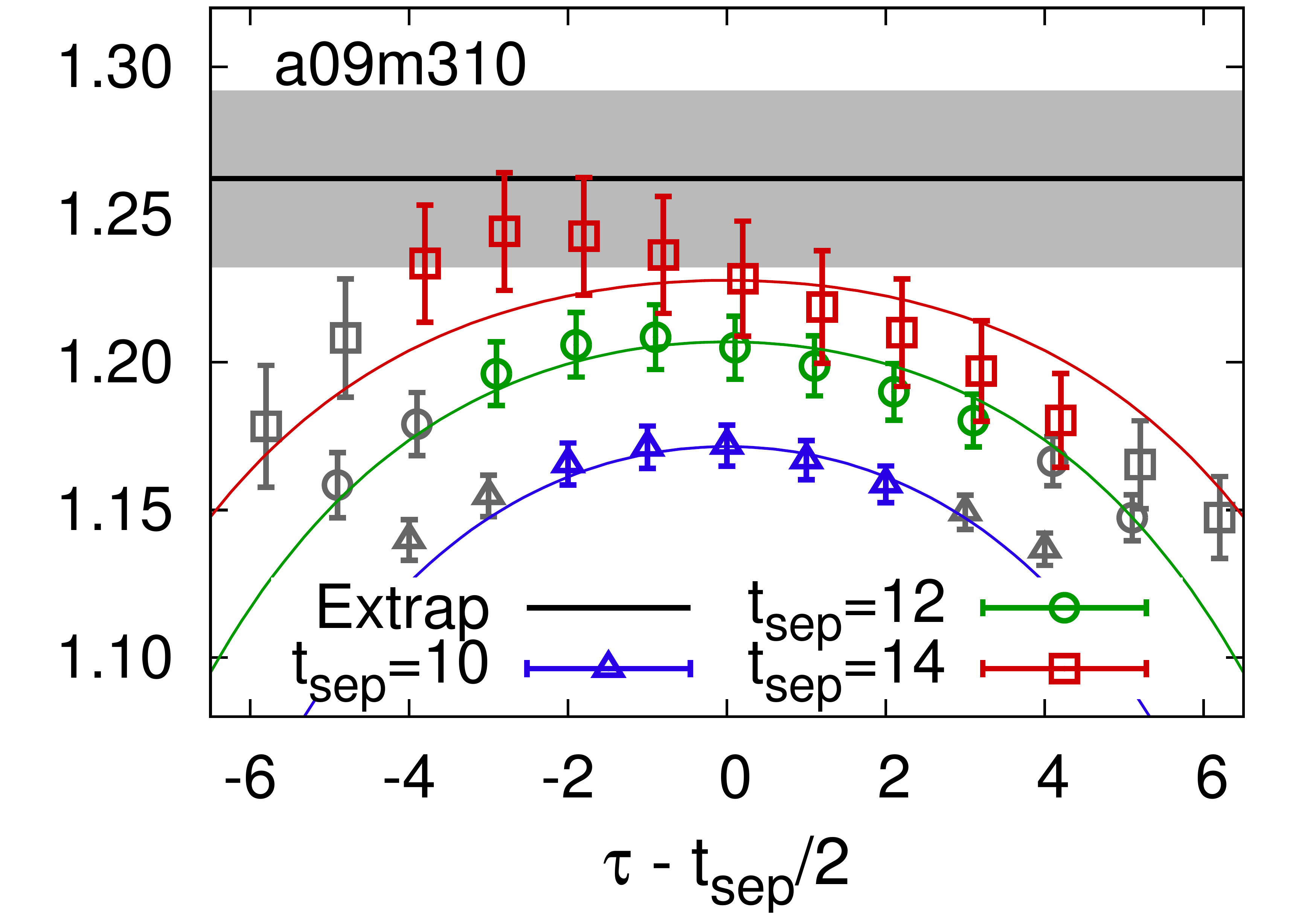}
\includegraphics[width=0.45\textwidth]{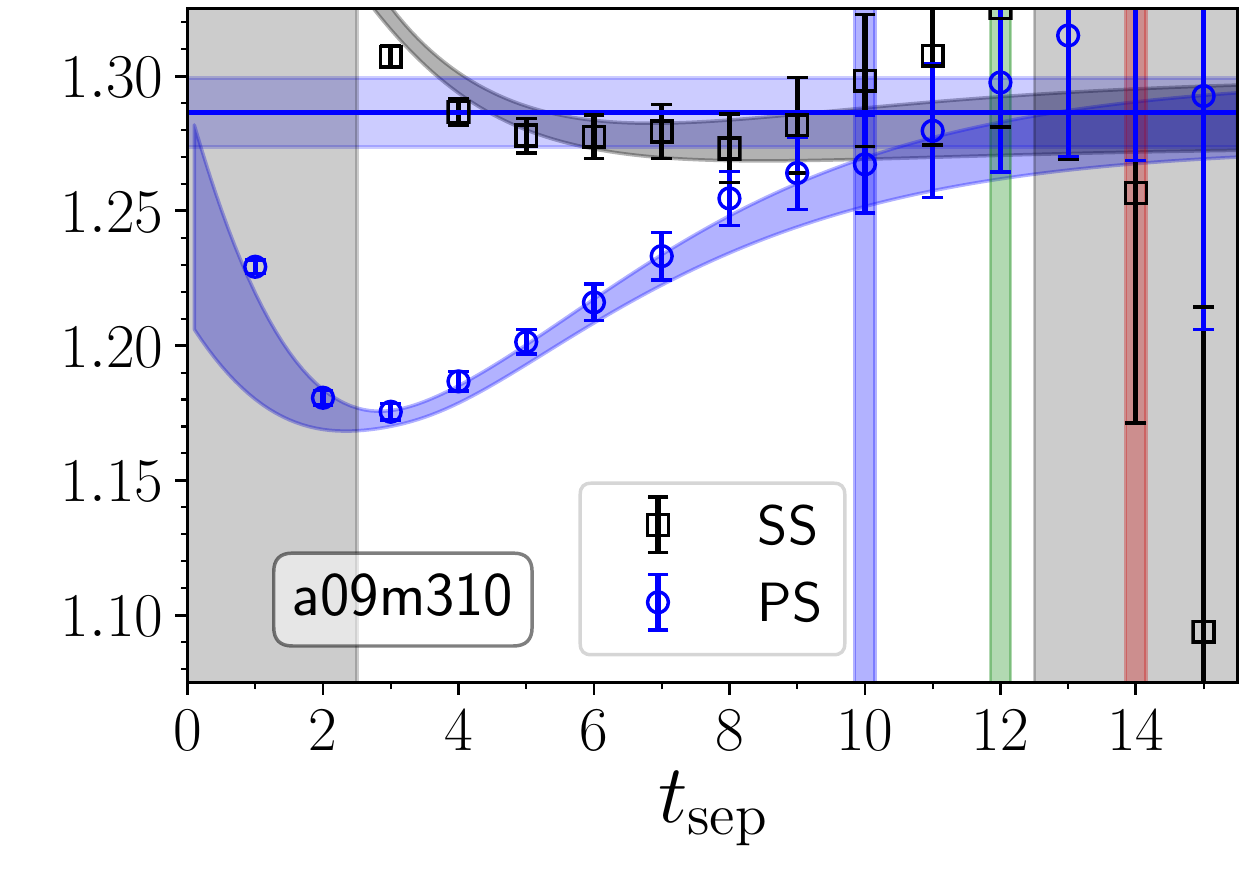}
\caption{\label{fig:fh_compare}
Left: Standard three-point function calculation from Ref.~\cite{Bhattacharya:2016zcn}.
Right: Sample fit from Ref.~\cite{Chang:2018uxx} on the same $a\sim0.09$~fm, $m_\pi\sim310$~MeV (a09m310) HISQ ensemble.
The vertical gray bands indicate results excluded from the fit.  The two sets of data correspond to two different sink smearings with a point (P) and smeared (S) sink and both use a smeared source.
The curves result from a simultaneous two-state fit to six correlation functions (SS and PS for the two-point, $g_V$ and $g_A$ correlation functions) while the horizontal band is the ground state matrix element.
The vertical bands correspond to the three values of $t_{\rm sep}$ used in Ref.~\cite{Bhattacharya:2016zcn}, $t_{\rm sep}=10$ (blue), $12$ (green), and $14$ (red).
The $y$-axis in both figures are the same.
As is evident, the FH method~\cite{Bouchard:2016heu} enables the use of many more values of $t_{\rm sep}$ as well as earlier values that are stochastically more precise and less prone to stochastic fluctuations, allowing for a more stable and precise extrapolation to $t_{\rm sep}\rightarrow \infty$.
}
\end{figure}

This unconventional method can be derived by applying the FHT to the effective mass of a correlation function in the presence of a background field.
Consider a two-point correlation function coupled to a external current
\begin{eqnarray}
C_\l(t) &=& \langle \O| \phi(t) \phi^\dagger(0) |\O\rangle_\l
\nonumber\\&=&
    \frac{1}{Z_\l} \int D[\phi] e^{-S}e^{-\l\int d^4 x j_\l(x)} \phi(t) \phi^\dagger(0),
    \quad\textrm{with}\quad
    Z_\l = \int D[\phi] e^{-S}e^{-\l\int d^4 x j_\l(x)}
\nonumber\\&=&
    \sum_n |\langle n| \phi^\dagger|\O\rangle_\l|^2 e^{-E_n^\l t}
\end{eqnarray}
The effective mass of this system will asymptote to the ground state energy for large $t$
\begin{equation}
    m_\lambda^{\rm eff}(t,\t) = \frac{1}{\t} \ln \left( \frac{C_\l(t)}{C_\l(t+\t)} \right)
    \mathop{\longrightarrow}\limits_{t\rightarrow\infty}
    E_0^\l\, .
\end{equation}
The Feynman-Hellmann Theorem in quantum mechanics relates matrix elements to shifts in the spectrum, $\partial_\l E_n^\l |_{\l=0} = \langle n| H_\l|n\rangle$.  If we ``follow our nose'' and apply the FHT to the effective mass, we derive a new correlation function that can be used to compute matrix elements in QFT~\cite{Bouchard:2016heu}
\begin{eqnarray}\label{eq:fh_meff}
\frac{\partial m_\l^{\rm eff}(t,\t)}{\partial \l} \bigg|_{\l=0}
    &=& \frac{1}{\t}\left[
        \frac{-\partial_\l C_\l(t+\t)}{C(t+\t)}
        -\frac{-\partial_\l C_\l(t)}{C(t)}
    \right]_{\l=0}
\nonumber\\&=&
    g_\l + z_{10} \left(
        \frac{e^{-(t+\t)\D_{10}} -e^{-t \D_{10}} }{\t}
    \right)+\cdots\, ,
\end{eqnarray}
where $g_\l$ is the matrix element of the ground state with the current, $g_\l = (2E_0)^{-1}\langle 0|j_\l|0\rangle,$
\begin{equation}\label{eq:C_lambda}
-\partial_\l C_\l(t) \Big|_{\l=0} =
    \frac{\partial_\l Z_\l}{Z}\bigg|_{\l=0} C(t)
    +\frac{1}{Z} \int D[\phi] e^{-S} \int d^3 x_\l dt_{\l}\
        O(t) j_\l(t_\l,x_\l) O^\dagger(0)\, ,
\end{equation}
is a new correlation function we can compute in the $\l=0$ vacuum, $C(t) = C_\l(t)|_{\l=0}$, $\D_{10} = E_1 - E_0$, $z_{10}$ is related to the ratio of the overlap of the interpolating operator onto the first excited state versus the ground state and the $\cdots$ represent contributions from higher excited states (see Ref.~\cite{Bouchard:2016heu} for a complete expression).
There are a few key features of this expression, \eq{eq:fh_meff}:
\begin{enumerate}
\item The excited state contributions are not only suppressed exponentially by the mass gap $\D_{10}$, but they are further suppressed by the difference between neighboring times separated by $\t$ (the user is free to chose $\t$ and we typically chose $\t=1$).  This enables a fit to the correlation function to begin earlier in Euclidean time than is generally possible for three-point correlation functions, where the stochastic signal is exponentially more precise;

\item Three point correlation functions depend upon two time variables, the source/sink separation time, $t$ (often denoted $t_{\rm sep}$) and the current insertion time, $t_\l$.  \eq{eq:fh_meff} only depends upon the source/sink separation time, simplifying the analysis;

\item In standard three point function calculations, $t=t_{\rm sep}$ is fixed and so multiple calculations must be performed to extrapolate to large $t$.  In the implementation of \eq{eq:fh_meff}, all values of $t$ are accessible with a single calculation, although for fixed current and momentum transfer;

\item Unlike other methods~\cite{Chambers:2014qaa,Chambers:2015bka,Savage:2016kon}, the variation of background field effects are not numerically implemented, but rather, the strength of the coupling to the background field, $\l$ is used to analytically track the dependence and derive this additional correlation function, \eq{eq:C_lambda} which is evaluated at $\l=0$.  Note, in the difference in \eq{eq:fh_meff}, the dependence upon the vacuum matrix element, the first term in \eq{eq:C_lambda}, exactly cancels.

\end{enumerate}
It is useful to compare and contrast results from the two methods, which we do in \fig{fig:fh_compare}.
A more exhaustive study of the sensitivity of the ground state matrix elements with respect to choices of $t_{min,max}$ on all correlation functions and all ensembles was performed and reported in Ref.~\cite{Chang:2018uxx}.
The final fits are very sensitive to the initial guess of all the fit parameters.
We therefore used a Bayesian fit to precondition the initial guess of a final two-state frequentist minimization, resulting in stable frequentist fits.  The bootstrap distributions of the ground state values are all close to Gaussian distributed with minimal tails, a further indication of the stability of the correlator fits.

\subsection{Extrapolation to the physical point \label{sec:extrap}}
In order to control the extrapolation to the physical point, several values of the lattice spacing, light quark masses and volumes must be used.
The only set of configurations that are publicly available with sufficient variation in these parameters are the $N_f=2+1+1$ HISQ~\cite{Follana:2006rc} ensembles generated by the MILC Collaboration~\cite{Bazavov:2012xda} which have been generated with six lattice spacings now spanning $0.03\lesssim a\lesssim0.15$~fm~\cite{Bazavov:2018omf} and three pion masses $m_\pi\sim\{130,220,310\}$~MeV.
In a previous study, we found that these three pion masses were not sufficient to eliminate the model dependence in the chiral extrapolation to the physical pion mass~\cite{Berkowitz:2017gql}.
Therefore, we generated six new ensembles with $a\sim\{0.09, 0.12, 0.15\}$~fm at $m_\pi\sim\{350,400\}$~MeV.  In total, we performed the calculation on 16 ensembles with $0.09 \lesssim a \lesssim 0.15$~fm, $130 \lesssim m_\pi \lesssim 400$~MeV and a dedicated volume study with three volumes on the a12m220 ensemble (we use the very convenient and descriptive shorthand notation for ensembles~\cite{Bhattacharya:2015wna}).

\begin{figure}[t]
\center
\includegraphics[width=0.55\textwidth]{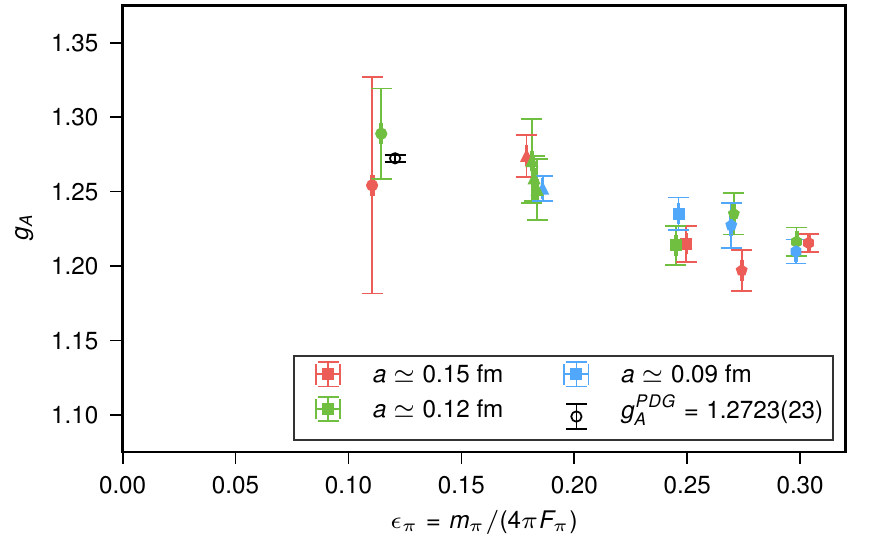}
\caption{\label{fig:ga_results}
Renormalized values of $g_A$ from Ref.~\cite{Chang:2018uxx}.
The results show a very mild dependence upon both $\e_\pi$ and $\e_a^2=a^2/(4\pi w_0^2)$, indicating a mild extrapolation.  While the results at the physical pion mass are significantly less precise than at heavier pion masses, the value on the a12m130 (green) ensemble has a 2.3\% uncertainty, which was the most precise value at the physical pion mass at the time of publication~\cite{Chang:2018uxx}.
}
\end{figure}

To perform the calculation, we first used gradient-flow~\cite{Narayanan:2006rf,Luscher:2011bx,Luscher:2013cpa} to smooth the UV fluctuations with a flow-time of $t_{gf}=1.0$ in lattice units.
We then solved M\"{o}bius Domain-Wall Fermions (MDWF)~\cite{Brower:2012vk} in the valence sector for an MDWF on gradient-flowed HISQ action~\cite{Berkowitz:2017opd}.
By holding the flow-time fixed in lattice units, any flow-time dependence should vanish as the continuum limit is taken.
With $t_{gf}=1$, we found that $m_{\rm res} \lesssim 0.1\times m_l$ for all ensembles with reasonable values of $L_5$, minimizing the residual chiral symmetry breaking.
Further, because of the near chiral symmetry of the action, we found that the non-perturbative value of $Z_A=Z_V$ to one part in $10^4$, greatly simplifying the renormalization of the matrix elements; the physical value of $g_A$ on each ensemble is simply given by $\mathring{g}_A / \mathring{g}_V$ where $\mathring{g}_{A,V}$ are the bare values of the charges, as $Z_V \mathring{g}_V =1$ by definition.
Our renormalized values of $g_A$ are shown in \fig{fig:ga_results}.

$g_A$ is a dimensionless quantity and so it is useful to construct dimensionless parameters that can be used to perform the three extrapolations to the physical point.
In order to perform the chiral extrapolation, we use the small parameter
\begin{equation}
\e_\pi = \frac{m_\pi}{4\pi F_\pi}\, ,
\end{equation}
where $F_\pi\simeq92$~MeV at the physical point.
This is also convenient as $\e_\pi$ is the small parameter which controls the chiral expansion in heavy-baryon $\chi$PT~\cite{Jenkins:1990jv}.
At fixed $m_\pi$, the finite volume corrections scale asymptotically in the volume as $e^{-m_\pi L}$~\cite{Luscher:1985dn} where $L$ is the size of the spatial box, for $m_\pi L \gtrsim 4$.
The leading finite volume corrections to $g_A$ were determined in Ref.~\cite{Beane:2004rf}.

To parameterize the continuum extrapolation, we introduce the small parameter
\begin{equation}
\e_a^2 = \frac{1}{4\pi}\frac{a^2}{w_0^2}\, ,
\end{equation}
where $w_0$ is a gradient-flow scale~\cite{Borsanyi:2012zs} which is $w_0\sim0.17$~fm.
For our MDWF on HISQ action, the leading discretization effects scale as $\e_a^2$ which follows from the Symanzik expansion of the lattice action~\cite{Symanzik:1983dc,Symanzik:1983gh} near the continuum limit.

At next-to-next-to-leading order (NNLO) in the chiral expansion, $g_A$ is given by~\cite{Kambor:1998pi}
\begin{equation}
g_A = g_0
    -\e_\pi^2 \left[ (g_0 + 2g_0^3) \ln(\e_\pi^2) -c_2 \right]
    +g_0 c_3 \e_\pi^3\, ,
\end{equation}
where $g_0$, $c_2$ and $c_3$ are low-energy-constants (LECs) that must be determined from analyzing LQCD results and/or experimentally measured observables.
The NLO expression (up to $\mathrm{O}(\e_\pi^2)$) was insufficient to describe the results for $m_\pi\lesssim310$~MeV and the NNLO expression has three LECs, so at least four values of the pion mass are required to fit the pion mass dependent LECs.

Given the very mild pion mass dependence observed in the results and the observed challenges with the convergence of baryon $\chi$PT~\cite{WalkerLoud:2008bp,WalkerLoud:2008pj,Walker-Loud:2013yua} we also explore a simple Taylor expansion both as an expansion about $\e_\pi^2$ and $\e_\pi$.  In each case, we explore the convergence of the expansions by performing each extrapolation with two different truncation orders.  For $\chi$PT, as the NNLO fit is the first to have a reasonable quality of fit, we add the counterterm from N$^3$LO to explore the convergence.  An honest full N$^3$LO fit is not possible as that contains 5 LECs and our results have 5 pion mass values (see \sec{sec:baryonChiPT}).
For the Taylor expansion fits, we perform both an NLO and NNLO fit where in each case, LO is just a constant in $\e_\pi$.
$\chi$PT fits with explicit delta-resonance degrees of freedom were not included as our numerical results do not include the $N\rightarrow\Delta$ and $\Delta\rightarrow\Delta$ matrix elements needed to constrain these contributions to $g_A$ and so too much prior knowledge from phenomenology would be required to stabilize the analysis with these states.
It should be noted that the large-$N_c$ expansion leads to a cancellation between virtual nucleon and delta contributions which can help explain the mild pion mass dependence~\cite{Hemmert:2003cb,CalleCordon:2012xz}.

The extrapolation analysis is performed in a Bayesian Framework allowing for a weighted model average to be performed.
By performing this semi-exhaustive chiral extrapolation analysis, we hope to remove theorist bias in the ``correct'' extrapolation form and let the numerical results dictate the preferred extrapolation.  Such a user unbiased approach is simple in this case in which the results have very mild pion mass dependence, but would be more challenging for quantities, such as the proton charge radius, which has a $\ln(m_\pi)$ divergence.
The resulting analysis is given in \tab{tab:model_selection}.
As can be seen, the Taylor expansion fits are strongly favored over the $\chi$PT fits.
In part, this is due to the strong cancellation among different orders in the $\chi$PT function that must occur to produce such a mild pion mass dependence.

\begin{table}[t]
\caption{\label{tab:model_selection}
Six chiral extrapolation models are considered.
For each fit, the \textit{augmented} $\chi^2/\textrm{dof}$, the \textit{log Gaussian Bayes Factor} (logGBF) $\mathcal{L}(D|M_k)$, the normalized weight of the fit determined from exp(logGBF) $P(M_k|D)$ and the resulting posterior at the physical point $P(g_A|M_k)$ is given.
The final uncertainty arises from the weighted average variance and the second is from the model variance, see Eq. (S26) of Ref.~\cite{Chang:2018uxx}.
}
\begin{center}
\footnotesize
\begin{tabular}{rcccc}
\hline\hline
Fit& $\chi^2/\textrm{dof}$& $\mathcal{L}(D|M_k)$& $P(M_k|D)$& $P(g_A|M_k)$\\
\hline
NNLO $\chi$PT               & 0.727 & 22.734& 0.033& 1.273(19) \\
NNLO+ct $\chi$PT          & 0.726 & 22.729& 0.033& 1.273(19) \\
NLO Taylor $\e_\pi^2$     & 0.792 & 24.887& 0.287& 1.266(09) \\
NNLO Taylor $\e_\pi^2$  & 0.787 & 24.897& 0.284& 1.267(10) \\
NLO Taylor $\e_\pi$        & 0.700 & 24.855& 0.191& 1.276(10) \\
NNLO Taylor $\e_\pi$     & 0.674 & 24.848& 0.172& 1.280(14) \\
\hline
\textbf{average}&&&& 1.271(11)(06) \\
\hline\hline
\end{tabular}
\end{center}
\end{table}

\begin{figure}[t]
\centering
\adjustbox{valign=t}{
\begin{subfigure}[t]{0.28\textwidth}
    \includegraphics[width=\textwidth]{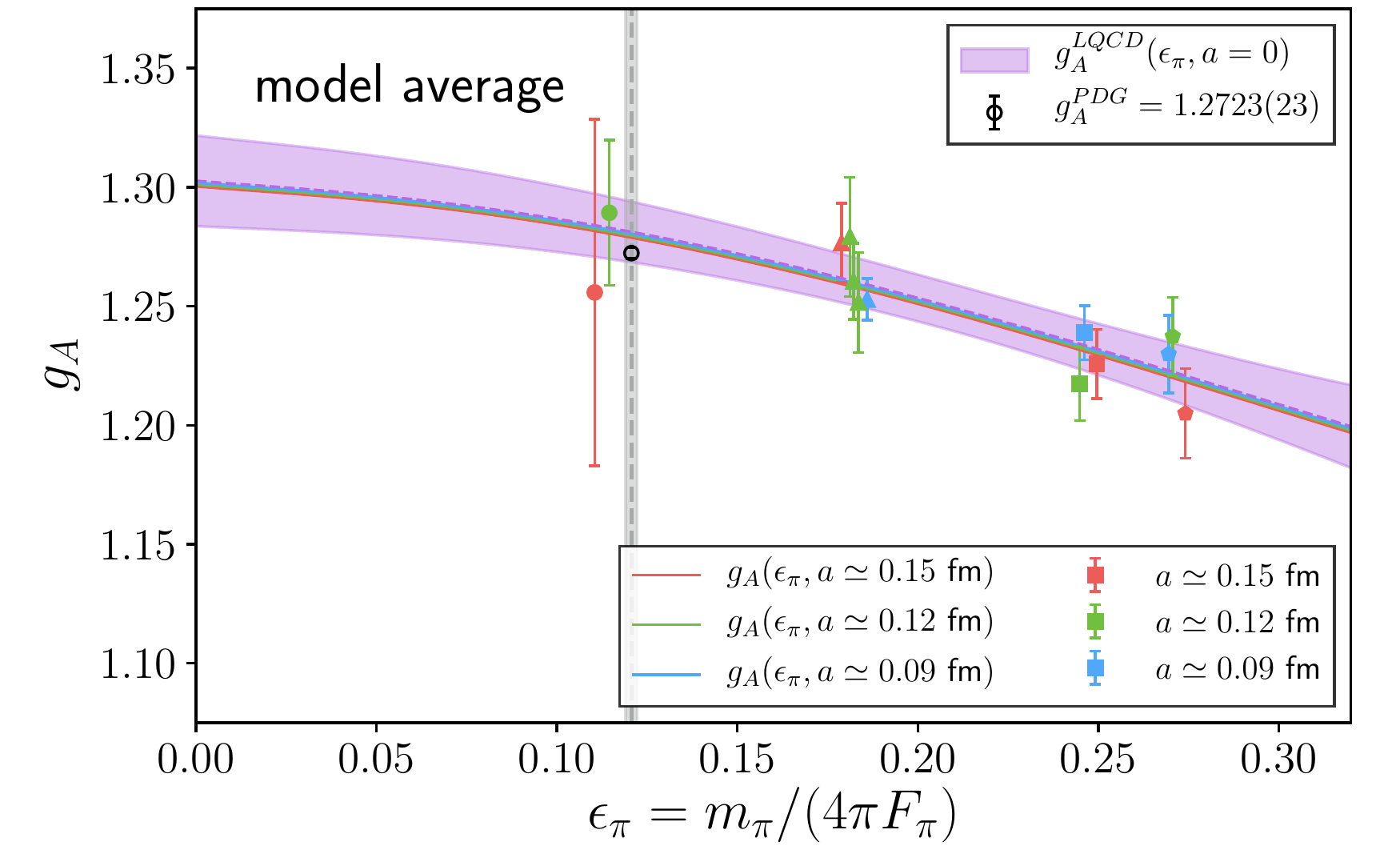}
    \caption{$m_\pi \lesssim 350$.}
    \includegraphics[width=\textwidth]{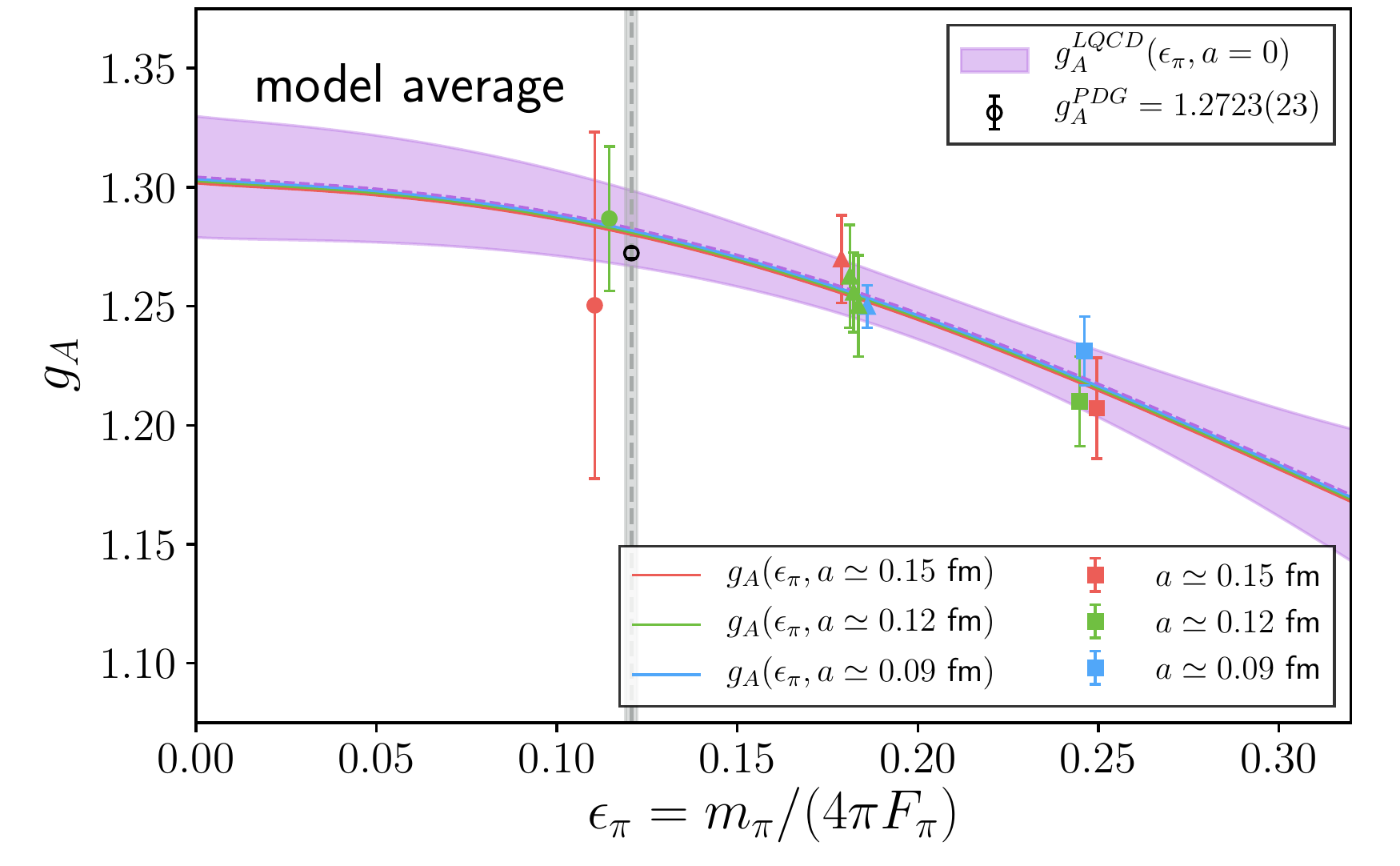}
    \caption{$m_\pi \lesssim 310$.}
\end{subfigure}
}\hspace{-0.6cm}
\adjustbox{valign=t}{
\begin{subfigure}[t][][r]{0.43\textwidth}
    \begin{center}
    \includegraphics[width=0.66\textwidth]{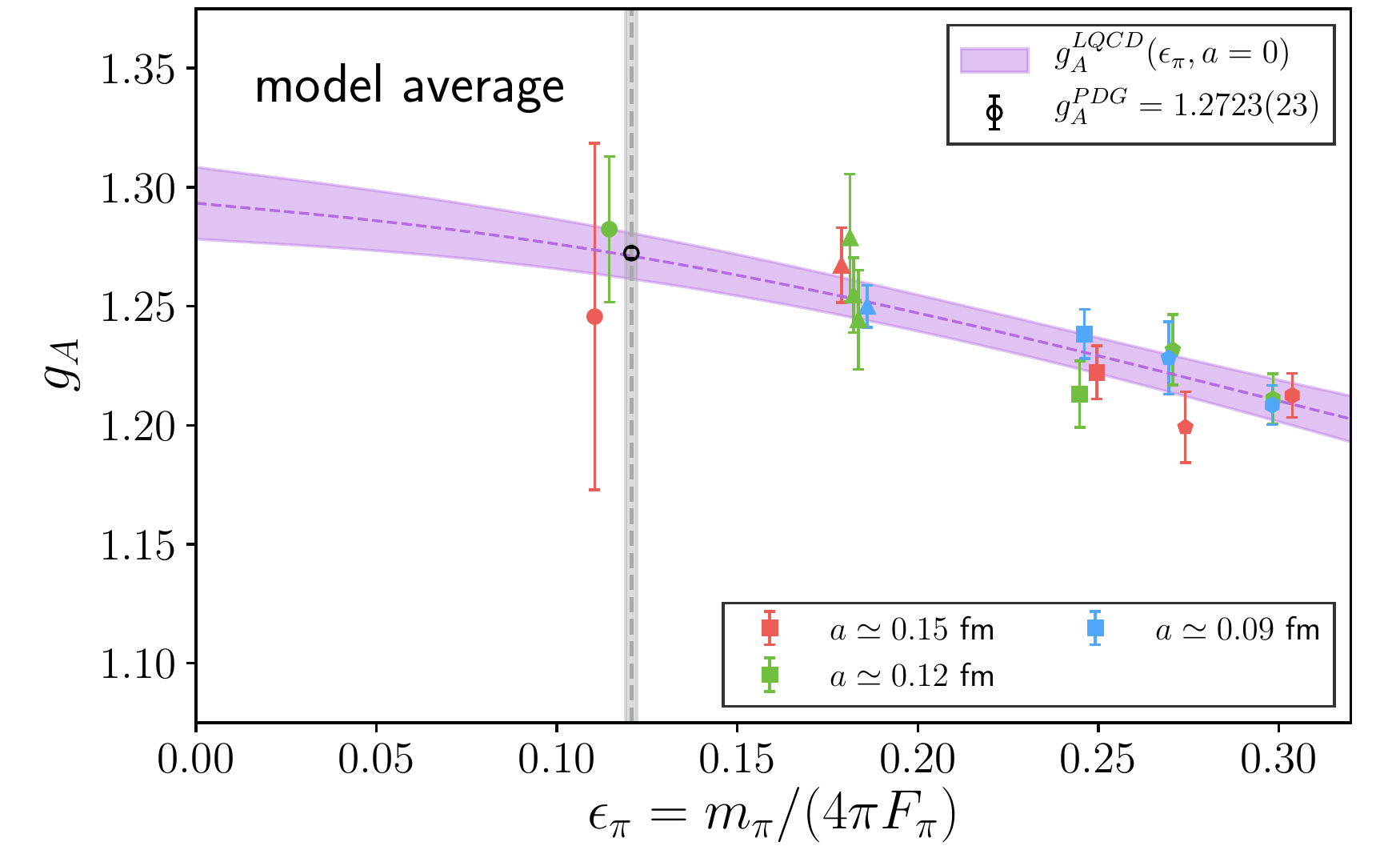}
    \end{center}
    \vspace{-0.5cm}\caption{All ensembles}
    \includegraphics[width=\textwidth]{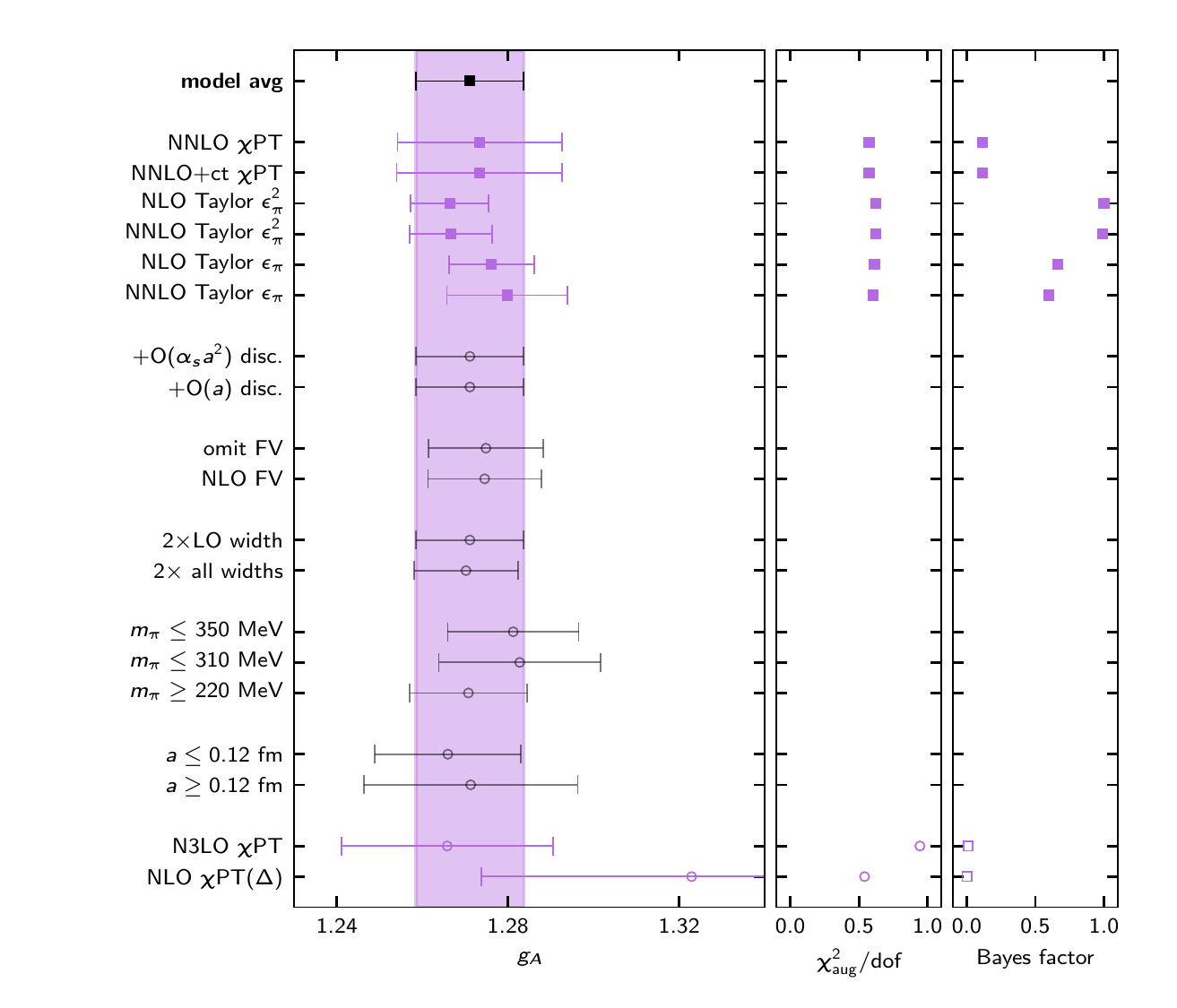}
\end{subfigure}
}\hspace{-0.6cm}
\adjustbox{valign=t}{
\begin{subfigure}[t]{0.28\textwidth}
    \includegraphics[width=\textwidth]{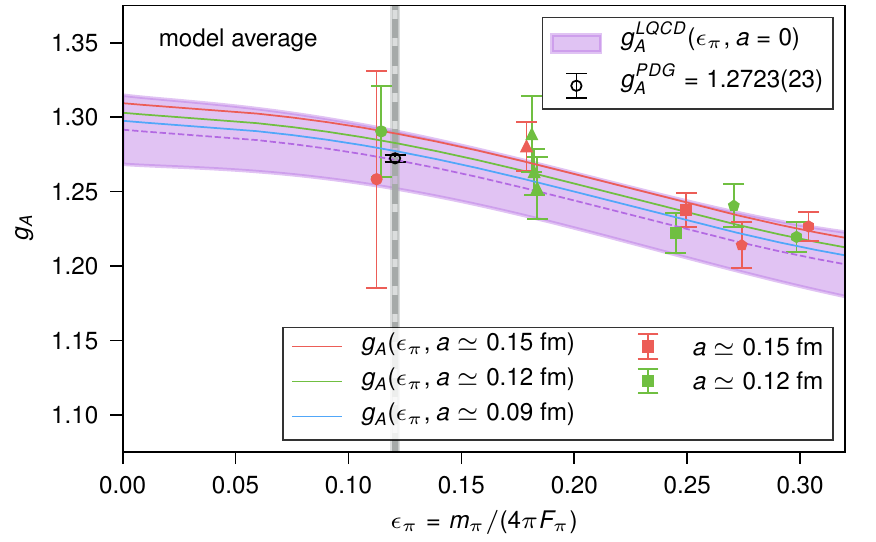}
    \caption{Remove a09 ensembles}
    \includegraphics[width=\textwidth]{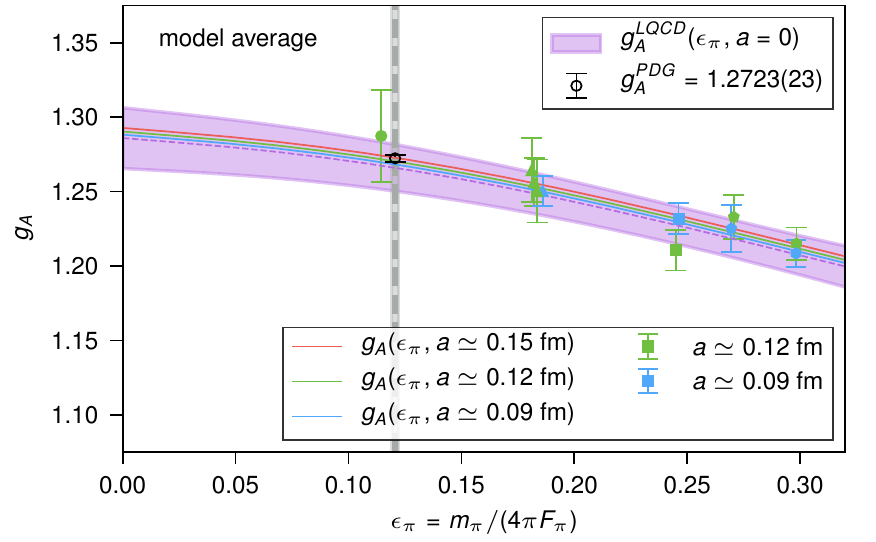}
    \caption{Remove a15 ensembles}
\end{subfigure}
}
\caption{\label{fig:stability}
Stability of the extrapolation as prior widths are varied, data is truncated and various continuum extrapolation models are included.  The magenta band in summary plot (bottom middle) is from the model average (top 6 entries) and displayed to guide the eye.  For all changes, the final extrapolation lies withing one standard deviation of the final answer, demonstrating the stability of the result.
}
\end{figure}

It is also important to check for the sensitivity of the final result upon the heavy pion mass points as these are the most statistically precise, but expected to have the largest systematic correction from the chiral extrapolation.
Such a study can be achieved either by adding more terms to the chiral extrapolation functions or by studying the result as the heavy mass points are cut.
The former method is already incorporated in our analysis in \tab{tab:model_selection}.
The latter method can not easily be used to quantitatively compare/average fits as fits that use different data sets can not easily be compared to each other as the absolute normalization, or \textit{evidence}, changes as data is added or removed.
Nevertheless, it is instructive to see how much the final extrapolation changes as data far from the physical point is removed.
We studied the sensitivity to data truncation by considering fits with $m_\pi \lesssim350$~MeV and $m_\pi \lesssim 310$~MeV as well as all data ($m_\pi \lesssim400$~MeV).
Further, we studied the extrapolation when we cut the $a\sim0.15$~fm or $a\sim0.09$~fm ensembles.
In addition to these data truncations, we studied the sensitivity to turning on/off the finite volume corrections, adding additional discretization corrections and increasing the size of the prior widths used to constrain the higher order corrections.
Results from this analysis are shown in \fig{fig:stability}.
As can be seen, the final extrapolated answer is very stable under all such variations/data truncations.

\subsection{Implications for $SU(2)$ baryon chiral perturbation theory \label{sec:baryonChiPT}}
\begin{figure}[t]\centering
\includegraphics[width=0.4\textwidth]{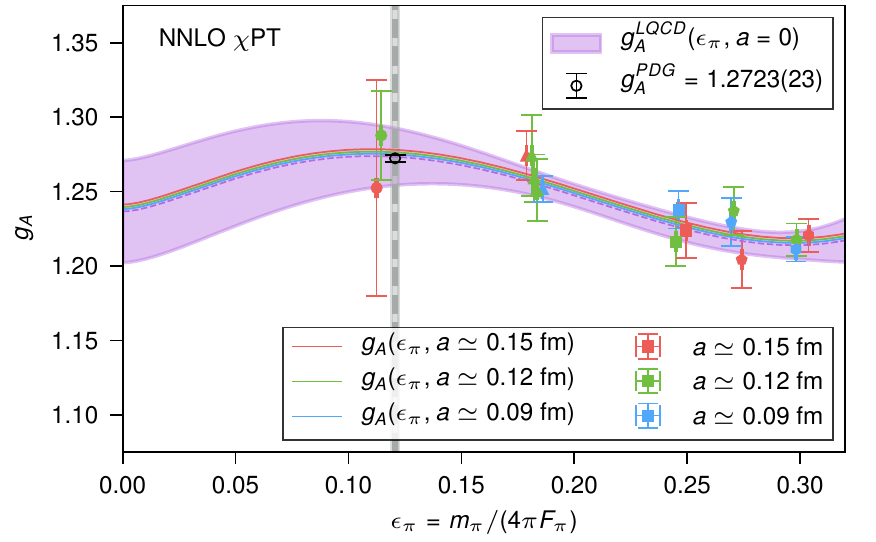}\quad
\includegraphics[width=0.4\textwidth]{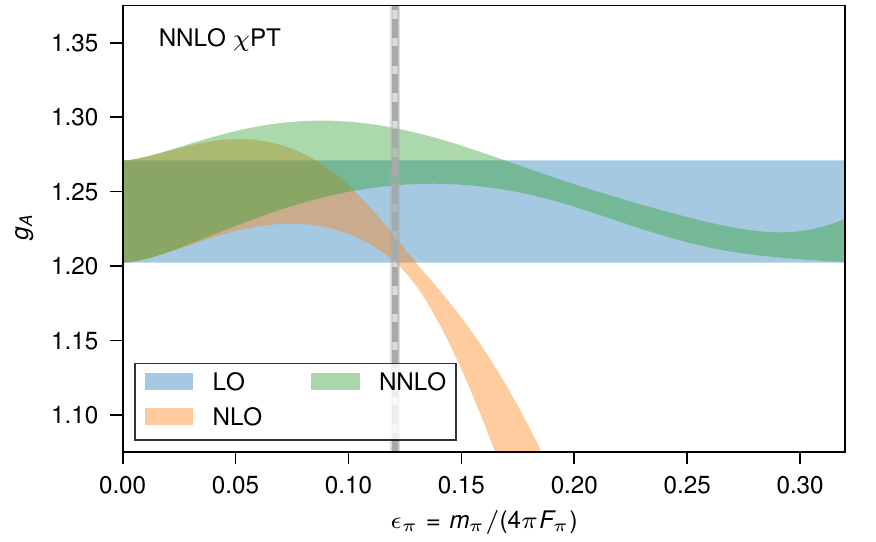}\\
\includegraphics[width=0.4\textwidth]{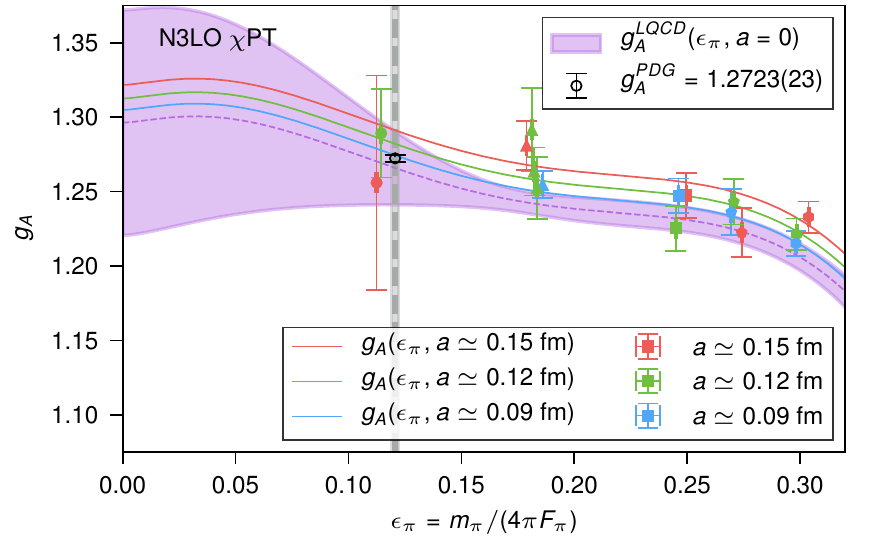}\quad
\includegraphics[width=0.4\textwidth]{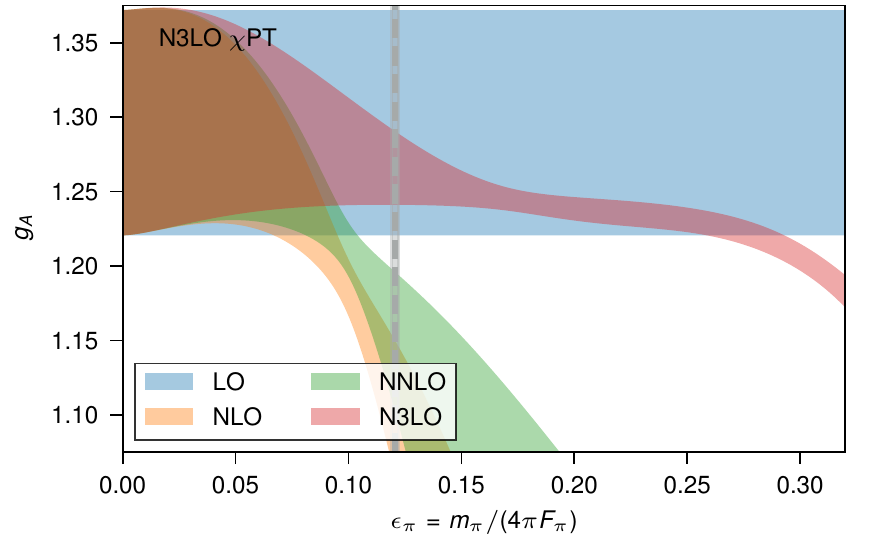}
\caption{\label{fig:gA_convergence}
Left: the NNLO and N$^3$LO extrapolation fits plotted versus $\e_\pi$.
Right: The cumulative contribution up to a given order versus $\e_\pi$.}
\end{figure}

As this is the Chiral Dynamics Workshop, I will spend some time discussing the implications for $SU(2)$ baryon $\chi$PT from these results.
The N$^3$LO extrapolation formula is known with the N$^3$LO terms given by (the $\ln^2$ coefficients differ from Ref.~\cite{Bernard:2006te} as we have converted $F\rightarrow F_\pi$)
\begin{equation}
\d g_A^\textrm{N$^3$LO} =
    \e_\pi^4 \left[ c_4
        +\tilde{\g}_4 \ln(\e_\pi^2)
        +\left(
            \frac{2}{3}g_0 +\frac{37}{12}g_0^3 +4 g_0^5
        \right) \ln^2(\e_\pi^2)
    \right]\, ,
\end{equation}
where $c_4$ and $\tilde{\g}_4$ are new LECs that must be constrained.
While an honest fit with the full N$^3$LO formula can not be performed (there are 5 LECs parameterizing the pion mass dependence and 5 values of $m_\pi$ in the numerical results), we can examine the stability of chiral extrapolation by performing this N$^3$LO analysis, which is shown in \fig{fig:gA_convergence}.
The right panels show the cumulative convergence of the fit.  For the NNLO analysis, one observes a rapid drop of the LO+NLO (NLO) contributions as $\e_\pi$ is increased slightly above its physical value.  At the physical pion mass, the NNLO contributions are opposite in sign and about twice as large as the NLO contributions.
Adding the N$^3$LO terms makes this situation worse: the NLO contribution dives faster and the convergence pattern worsens.  At the physical pion mass, we find the order-by-order contributions
\begin{equation}
\begin{array}{c|cccc}
\textrm{N$^n$LO}& \d g_A^\textrm{LO}& \d g_A^\textrm{NLO}& \d g_A^\textrm{N$^2$LO}& \d g_A^\textrm{N$^3$LO}\\
\hline
\textrm{N$^2$LO}& 1.237(34)& -0.026(30)& 0.062(14)& -\\
\textrm{N$^3$LO}& 1.296(76)& -0.19(12) & 0.045(63)& 0.117(66)\\
\hline
\end{array}\, ,
\end{equation}
for which there are strong cancellations order-by-order for $SU(2)$ heavy-baryon $\chi$PT.
It is not expected the covariant formulation of baryon $\chi$PT~\cite{Becher:1999he} will improve the situation but the inclusion of explicit delta-degrees of freedom should.
However, in the case of the nucleon mass, the virtual delta-corrections add with the same sign and so they will make the convergence pattern of the nucleon mass worse.  All in all, lattice results are indicating that $SU(2)$ heavy-baryon $\chi$PT without delta-degrees of freedom is a failing perturbative expansion, even at the physical pion mass.

\section{Updates and Outlook \label{sec:future}}
Improving the precision of $g_A$ from QCD is interesting for several reasons.  Already, the precision of $g_A$~\cite{Chang:2018uxx} sets the limiting constraint on right-handed BSM currents~\cite{Alioli:2017ces} and further improvements are welcome.
A reduction of the uncertainty to 0.2\% would place an uncertainty on the predicted neutron lifetime, using $g_A$ from QCD, at a level sufficient to provide 4-sigma discriminating resolution.
This is a tractable problem for LQCD in the exascale computing era.  However, to reduce the uncertainty below 0.5\% requires a comparison with the newly uncovered QED radiative corrections to $\beta$-decay~\cite{Hayen:2019nic} (there is also an improved determination of the inner radiative corrections~\cite{Seng:2018yzq,Seng:2018qru} which can be tested with LQCD as well~\cite{Seng:2019plg}).

\begin{figure}[t]\centering
\includegraphics[width=0.45\textwidth]{figures/gA_nature}\quad
\includegraphics[width=0.45\textwidth]{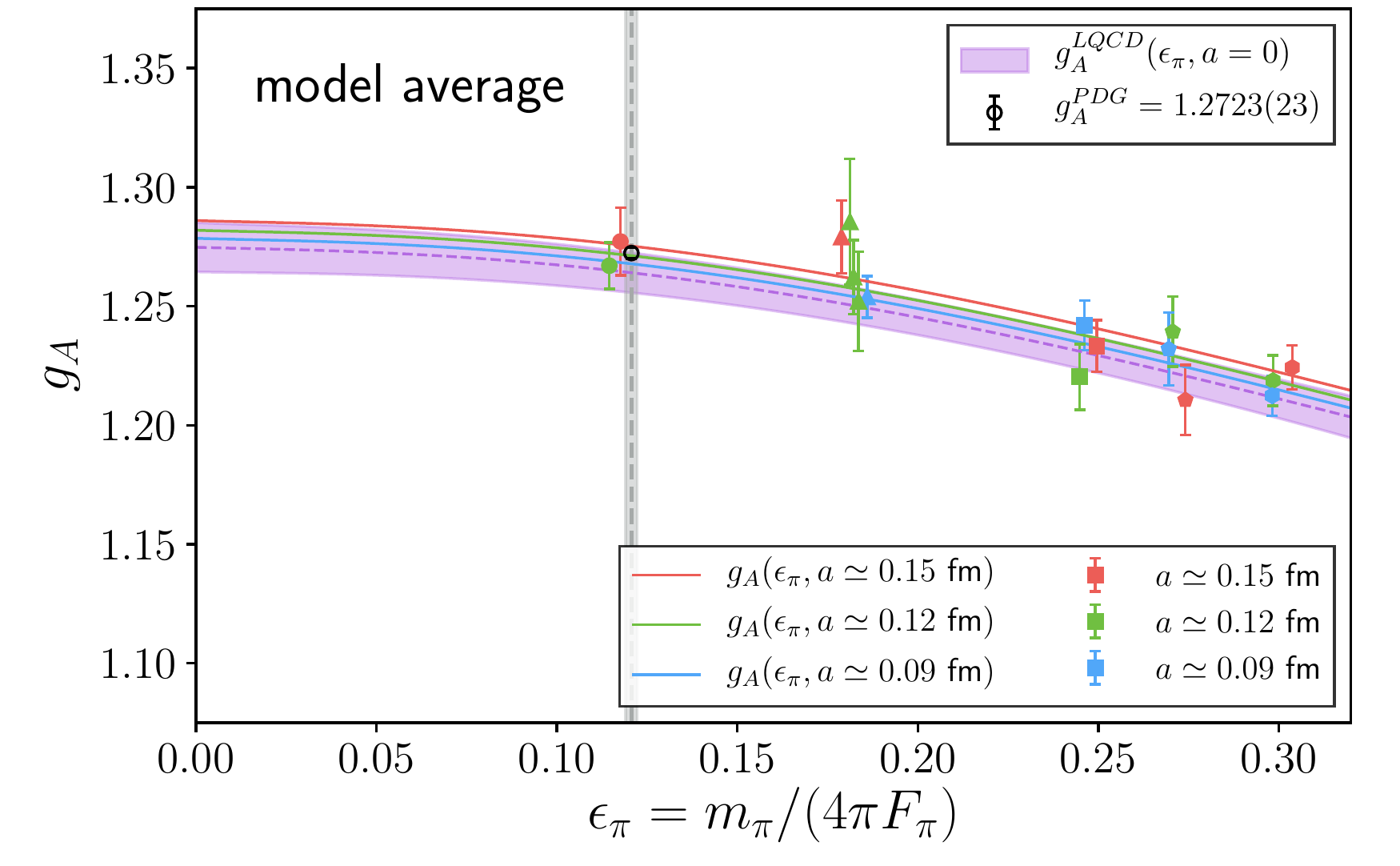}
\caption{\label{fig:gA_update}
Left: our published model-average extrapolation~\cite{Chang:2018uxx}.
Right: A preliminary update of our results with improved statistics at the physical pion mass enabled through early science time on Sierra at LLNL.}
\end{figure}

In the current near-exascale era, we will see significant improvements.
Machine-to-machine, for our LQCD applications, Summit is 15 times faster than Titan~\cite{Berkowitz:2018gqe}.
We are continuing to improve our determination of $g_A$ as part of a more comprehensive program to determine the nucleon elastic form factors.
We were early-science users on the Sierra Supercomputer at LLNL in late 2018.
In \fig{fig:gA_update}, we show our preliminary updated results with this early science time (right) with our published result~\cite{Chang:2018uxx} (left).
Of note
\begin{enumerate}
\item The a12m130 ensemble (left most green point) is the most expensive one used in this work.  In our published result, we had three sources per configuration, which cost more than all other ensembles combined.  These were produced with our 2016 INCITE allocation on Titan.
In 2.5 weekends on Sierra, we were able to produce 16 sources per configuration, and the updated result now has 32 sources, can be fit with an unconstrained 3-state frequentist fit, and the precision on this ensemble is sub-percent;

\item The a15m130 ensemble (left most red point) was too noisy for this project, likely from the small volume ($L=32, T=48$ with $m_\pi L\sim3.23$).  We generated a new ensemble, a15m135XL ($L=48$, $T=64$) with 4 streams of 250 configurations each.  The right panel of \fig{fig:gA_update} has a result from this ensemble with 8 sources per configuration.

\item Our preliminary update with new a12m130 and a15m135XL results has a 0.74\% uncertainty
\begin{equation}
g_A^{\rm QCD} = 1.2711(125) \rightarrow 1.2642(93)\, .
\end{equation}

\item We have generated new ensembles at $m_\pi\sim\{180,260\}$~MeV to increase the density of light pion mass results, and help explore the convergence of baryon $\chi$PT.

\item To achieve a 0.5\% uncertainty, a fourth lattice spacing at $a\sim0.06$~fm is likely required with our MDWF on gradient-flowed HISQ action~\cite{Berkowitz:2017opd}.

\end{enumerate}

\bigskip\noindent\textbf{Acknowledgments}
We thank the LLNL Multiprogrammatic and Institutional Computing program for Grand Challenge allocations on the LLNL supercomputers, Surface, RZHasGPU and Vulcan. This research also used the NVIDIA GPU-accelerated Titan supercomputer at the Oak Ridge Leadership Computing Facility at the Oak Ridge National Laboratory, which is supported by the Office of Science of the U.S. Department of Energy under Contract No. DE-AC05-00OR22725, through an award of computer time provided by the INCITE program.
This research also used the Sierra computer operated by the Lawrence Livermore National Laboratory for the Office of Advanced Simulation and Computing and Institutional Research and Development, NNSA Defense Programs within the U.S. Department of Energy, during the Early Science period.

\bibliographystyle{JHEP}
\bibliography{ga_bib}

\end{document}